\documentclass{article}
\usepackage{spconf,amsmath,graphicx,hyperref}
\usepackage[subtle]{savetrees}
\usepackage{siunitx}
\usepackage{booktabs}
\usepackage{multirow}

\title{DFKI-Speech System for WildSpoof Challenge:\\ A robust framework for SASV In-the-Wild}
\name{
Arnab Das$^{1,2}$,
Yassine El Kheir$^{1,2}$,
Enes Erdem Erdogan$^{1,3}$,
Feidi Kallel$^{1,3}$,
Tim Polzehl$^{1,2}$,
Sebastian M{\"o}ller$^{1,3}$
}
\address{
  $^1$German Research Center for Artificial Intelligence (DFKI), Berlin, Germany\\
  $^2$GretchenAI, Berlin, Germany\\
  $^3$Technical University of Berlin, Berlin, Germany\\
  \small \texttt{arnab.das@dfki.de}
}
\begin{document}
%\ninept
%
\maketitle
\begin{abstract}
This paper presents the DFKI-Speech system developed for the WildSpoof Challenge under the Spoofing-aware Automatic Speaker Verification (SASV) track. 
We propose a robust SASV framework in which a spoofing detector and a speaker verification (SV) network operate in tandem.
The spoofing detector employs a self-supervised speech embedding extractor as the frontend, combined with a state-of-the-art graph neural network backend.
In addition, a top-3 layer–based mixture-of-experts (MoE) is used to fuse high-level and low-level features for effective spoofed utterance detection.
For speaker verification, we adapt a low-complexity convolutional neural network that fuses 2D and 1D features at multiple scales, trained with the SphereFace loss.
Additionally, contrastive circle loss is applied to adaptively weight positive and negative pairs within each training batch, enabling the network to better distinguish between hard and easy sample pairs.
Finally, fixed imposter cohort–based AS-Norm score normalization and model ensembling are used to further enhance the discriminative capability of the speaker verification system. 
\end{abstract}
\begin{keywords}
Speaker-verification, anti-spoofing, SASV in-the-wild, fine-tuning, WildSpoof
\end{keywords}
\section{Introduction}
\label{sec:intro}
With the rise of spoofed speech attacks, safeguarding automatic speaker verification systems has become an urgent necessity.
To develop a secure and robust SASV system, a foremost requirement is a dataset that reflects real-world variability and challenges.
However, existing datasets often comprise clean utterances recorded in controlled laboratory environments, failing to capture the complexities of in-the-wild data.
In contrast, the SpoofCeleb dataset \cite{jung2025spoofceleb} is derived from the VoxCeleb1 corpus and comprises speech collected under diverse, noisy real-world conditions.
As part of the WildSpoof Challenge, we develop a robust and accurate SASV system that demonstrates substantially superior performance compared to existing baselines, such as SKA-TDNN \cite{mun2023frequency}.

\section{System Description}
\label{sec:sys_desc}
\begin{figure}[ht]
    \centering
    \includegraphics[width=\columnwidth]{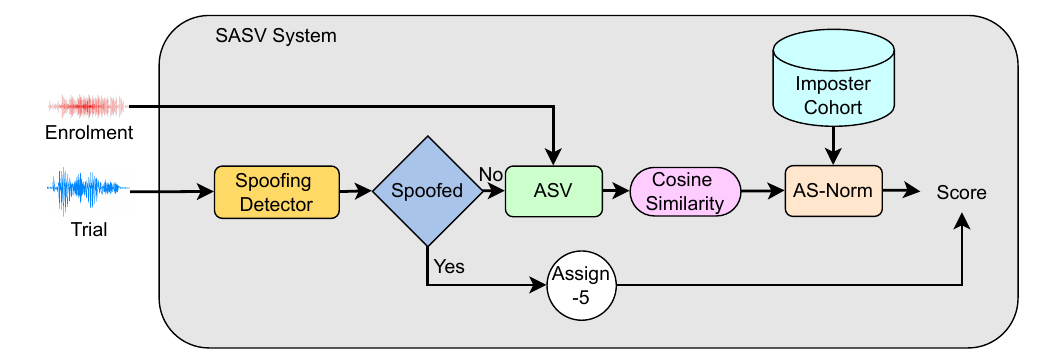}
    \caption{Proposed SASV framework architecture.}
    \label{fig:architecture}
\end{figure}

The high-level architecture of the proposed SASV framework is illustrated in Fig. \ref{fig:architecture}. 
A trial utterance is first processed by the spoofing detector (SD), which performs binary classification to determine whether the input is spoofed.
If the utterance is classified as spoofed, it is immediately rejected and assigned a score of $-5$.
Otherwise, the trial is forwarded to the ASV network.
The ASV module extracts speaker embeddings for both the trial and enrollment utterances, after which cosine similarity is computed between the embeddings.
This similarity score is then normalized using top-300 AS-Norm \cite{yin2008adaptive, yakovlev2024reshape} with a pre-selected imposter cohort to further enhance its discriminative capability. 
Finally, based on the normalized score, the trial is accepted or rejected.
\newline
\textbf{A. Spoofing detector:} The high-level architecture of the spoofing detector network we adopt can be found it \cite{wang2025mixture}.
As the frontend, we employ a pre-trained self-supervised wav2vec 2.0 XLS-R model from the fairseq repository \cite{babu2021xlsr}, which consists of initial convolutional layers followed by 24 transformer layers. 
This model produces frame-wise speech embeddings with a dimensionality of 1024 per frame.
To enhance spoofing detection performance, we incorporate both low-level and high-level feature fusion from different transformer layers using a sparse top-3 \textbf{mixture-of-experts (MoE)} mechanism as recommended by \cite{wang2025mixture}.
The embedding from the final transformer layer is fed into a gating module to generate a softmax distribution over the remaining layers. Only the top three layers are dynamically selected, while the others are zeroed out.
The embeddings from the selected layers are then summed with the final-layer embedding and passed to the backend.
For the backend, we adopt the graph neural network–based AASIST architecture.
Unlike \cite{wang2025mixture}, both the front and back ends are jointly trained. 
We train the network with binary cross-entropy loss. 
\newline
\textbf{B. Automatic speaker verification:} For the ASV task, we adopt the low-complexity Reshape Dimensions Network (ReDimNet) architecture with 15M parameters, as proposed in \cite{yakovlev2024reshape}.
The network takes mel-spectrograms as input.
ReDimNet comprises multiple transformer-based self-attention temporal encoder blocks that produce 1D feature maps, along with several ConvNeXt blocks that generate 2D feature maps at multiple scales and resolutions.
To enhance utterance-level speaker-discriminative feature extraction, feature mixing is performed via multiscale residual connections between the 1D and 2D representations.
The network is trained using the SphereFace loss \cite{wen2021sphereface2} with a PK batch sampling strategy \cite{hermans2017defense}, where each batch comprises K samples from each of P speakers.
In addition, contrastive Circle loss \cite{sun2020circle} is employed to further improve inter-speaker discrimination and intra-speaker compactness.
ReDimNet produces speaker embeddings with a dimensionality of 192.
\textbf{i) SphereFace loss:} SphereFace \cite{yakovlev2024reshape,wen2021sphereface2} enforces multiplicative angular margins in the hyperspherical embedding space, in contrast to additive angular margin, promoting compact intra-class distributions and large inter-class angular separation as depicted in Eq. \ref{eq:sface}.
Here $y_i$ is the ground-truth speaker label of sample $i$, $\theta_{i,j} = arccos(w_j^Tx_i)$ is the angle between embedding and class weight and $m,s$ are the margin and scale factor respectively.
By replacing the target logit with a margin-augmented angular function, it constrains the decision boundaries on the unit hypersphere, thereby enhancing the discriminative power of speaker embeddings.
\begin{equation}
\label{eq:sface}
\mathcal{L}_{\text{SphereFace}}
= - \log
\frac{
e^{s \cos(m \theta_{i,y_i})}
}{
e^{s \cos(m \theta_{i,y_i})}
+ \sum\limits_{j \neq y_i} e^{s \cos(\theta_{i,j})}
}
\end{equation} 
\textbf{ii) Circle loss:} Circle Loss \cite{sun2020circle} formulates metric learning as an optimization problem focused on pairwise similarity, in which positive and negative pairs are adaptively weighted according to their similarity.
By emphasizing hard positives and hard negatives while suppressing easy pairs, it encourages a well-structured and highly discriminative embedding geometry.

\begin{equation}
\label{eq:circle}
\mathcal{L}_{\text{Circle}}
=
\log \left(
1
+
\sum\limits_{p}
e^{-\gamma \alpha_p (s_p - \Delta_p)}
\sum\limits_{n}
e^{\gamma \alpha_n (s_n - \Delta_n)}
\right)
\end{equation}

\textbf{iii) AS-Norm:} Adaptive s-normalization (AS-Norm) \cite{yin2008adaptive, yakovlev2024reshape} is a post-processing technique that reduces channel and domain mismatch in speaker verification by normalizing similarity scores using cohort statistics. It symmetrically applies score normalization to both the enrollment and trial utterances, thus improving the robustness and calibration of verification scores.
This is depicted in Eq. \ref{eq:asnorm}, where $s(e,t)$ is the cosine similarity between enrollment embedding $e$ and trial embedding $t$ and $\mu, \sigma$ are the mean and standard deviation of scores between cohort embeddings with either enrolment $e$ or trial $t$. 
\begin{equation}
\label{eq:asnorm}
s_{\text{AS-Norm}}
=
\frac{1}{2}
\left(
\frac{s(e,t) - \mu_e}{\sigma_e}
+
\frac{s(e,t) - \mu_t}{\sigma_t}
\right)
\end{equation}

\section{Experiment Setup}
\label{sec:exp_setup}
\textbf{Datasets:}
We train both the spoofing detector and the ASV model using the official SpoofCeleb dataset \cite{jung2025spoofceleb}, which comprises bona fide utterances from more than 1,250 speakers and spoofed utterances generated using 23 diverse TTS systems.
SpoofCeleb is based on speech from the VoxCeleb1 corpus and reflects real-world, diverse, and noisy recording conditions, in contrast to existing controlled datasets. 
All utterances are resampled to \SI{16}{\kilo\hertz} and either padded or truncated to a duration of 4 seconds.
During ASV system training, we apply data augmentation by adding random noise from the MUSAN dataset, convolving with room impulse responses from the RIR dataset, and applying RawBoost augmentation along with random gain.
\newline \textbf{Implementation details:}
The models were trained on H100 GPU using the Adam optimizer with a learning rate of $10^{-6}$ for the spoofing detection models and $10^{-4}$ for the ASV model.
We initialize the ASV model with pretrained weights from the b6 model available at \nolinkurl{https://github.com/IDRnD/redimnet/blob/master/EVALUATION.md}, which was trained on the VoxCeleb2 dataset.
For the Sphere-Face loss, we set the scale parameter (s = 30) and the angular margin (m = 1.5), while the weight for the Circle loss is set to 0.2.

\begin{table}
\centering
\caption{SASV performance on SpoofCeleb eval. set}
\label{tab:eval_results}
\small
\begin{tabular}{l|lll}
\toprule
Systems             & a-DCF   & SV-EER & SPF-EER  \\
\midrule
\multicolumn{4}{l}{Baselines} \\
\midrule
Conventional ASV \cite{jung2025spoofceleb}   & 0.4923  & 3.84   & 23.44    \\
SASV - OOD Data \cite{jung2025spoofceleb}  & 0.9998  & 38.94  & 52.24    \\
SASV - SpoofCeleb  \cite{jung2025spoofceleb} & 0.2902  & 12.78  & 5.00  \\
\midrule
\multicolumn{4}{l}{Ours}                          \\
\midrule
SD+ASV-Pretrained   & 0.036   & 2.74 & 0.01   \\
SD+ASV-SpoofCeleb & 0.04    & 3.53 & 0.01   \\
Ensemble            & \textbf{0.03183} & \textbf{2.45} & \textbf{0.01}  \\
\bottomrule
\end{tabular}
\end{table}

\section{Results}
\label{sec:results}
We first evaluate our system on the SpoofCeleb evaluation set and report a-DCF, SV-EER, and SPF-EER for spoofing, as recommended in \cite{jung2025spoofceleb}, together with three baseline results.
The results are presented in Tab. \ref{tab:eval_results}, which shows substantial improvement compared to the baselines.
Our proposed SASV framework achieves an a-DCF score of 0.036 when using the pretrained ASV model, which slightly increases to 0.04 after fine-tuning on the SpoofCeleb dataset. The best performance is obtained through model ensembling, yielding an a-DCF of approximately 0.032 and an SV-EER of 2.45%.

The performance of our framework on the test data provided by the WildSpoof challenge is summarized in the Tab. \ref{tab:test} as well as two baseline systems.
Our proposed framework achieves an overall macro a-DCF score of $0.2022$. 
For the WildSpoof and SpoofCeleb subsets, the scores are $0.1508$ and $0.0406$, respectively, demonstrating a substantial improvement over the baseline.

\begin{table}
\centering
\caption{SASV performance on challenge test subsets.WSpoof is the WildSpoof TTS systems.}
\label{tab:test}
\small
\begin{tabular}{@{}l|lllll@{}}
\toprule
System & a-DCF                                         & WSpoof  & SpoofCeleb    & ASV5     & ASV22         \\ 
\hline
B01    & \textcolor[rgb]{0.318,0.318,0.318}{0.3715}          & \textcolor[rgb]{0.318,0.318,0.318}{0.3818}          & \textcolor[rgb]{0.318,0.318,0.318}{0.1677}          & \textcolor[rgb]{0.318,0.318,0.318}{0.6610}          & \textcolor[rgb]{0.318,0.318,0.318}{0.4689}           \\ 
\hline
B02    & \textcolor[rgb]{0.318,0.318,0.318}{0.7513}          & \textcolor[rgb]{0.318,0.318,0.318}{0.8677}          & \textcolor[rgb]{0.318,0.318,0.318}{0.8791}          & \textcolor[rgb]{0.318,0.318,0.318}{0.5460}          & \textcolor[rgb]{0.318,0.318,0.318}{0.4682}           \\ 
\hline
Ours [T04]   & \textcolor[rgb]{0.318,0.318,0.318}{\textbf{0.2022}} & \textcolor[rgb]{0.318,0.318,0.318}{\textbf{0.1508}} & \textcolor[rgb]{0.318,0.318,0.318}{\textbf{0.0406}} & \textcolor[rgb]{0.318,0.318,0.318}{\textbf{0.4882}} & \textcolor[rgb]{0.318,0.318,0.318}{\textbf{0.3420}}  \\
\hline
\end{tabular}
\end{table}

\section{Conclusion}
\label{sec:conclu}
In this paper, we present the DFKI-Speech group’s SASV solution for the WildSpoof Challenge. Our proposed framework integrates a dedicated spoofing countermeasure with a speaker verification system. Evaluation results demonstrate that the proposed approach effectively learns the SASV task with a high degree of accuracy and substantially outperforms the baseline.

\bibliographystyle{IEEEbib}
\bibliography{strings,refs}

\end{document}